\documentclass[final,3p,times]{elsarticle}

\usepackage{natbib}

 \usepackage{setspace}

 \usepackage{boondox-calo} 

\usepackage[makeroom]{cancel}

\renewcommand\refname{References}

\usepackage[normalem]{ulem}	

\usepackage{subfigure}	

\newcounter{lastnote}

\usepackage{amsthm,amsmath}
\RequirePackage[OT1]{fontenc}
\RequirePackage{amsthm,amsmath}

\RequirePackage[colorlinks,citecolor=blue,urlcolor=blue]{hyperref}
\usepackage[latin1] {inputenc}
\usepackage{vmargin}
\usepackage{multirow}
\usepackage{bbm}

\usepackage{amsthm, dsfont, tipa}
\usepackage{graphics}
\usepackage{graphicx,float,multirow,subfigure}

\usepackage{tikz} 

\usetikzlibrary{arrows,decorations.markings,chains,shapes.arrows,shapes.misc,fit, shapes}

 \tikzset{
  big arrow/.style={
    decoration={markings,mark=at position 1 with {\arrow[scale=2,#1]{>}}},
    postaction={decorate},
    shorten >=0.4pt},
  big arrow/.default=black}

\theoremstyle{plain}

\newtheorem{lemma}{Lemma}

\usepackage{amssymb}
\usepackage{epsfig}

\newcommand{\LR}{\textrm{LR}}

\newcommand{\appropto}{\mathrel{\vcenter{
  \offinterlineskip\halign{\hfil$##$\cr
    \propto\cr\noalign{\kern2pt}\sim\cr\noalign{\kern-2pt}}}}}

\usepackage{lineno}
\begin{document}


\begin{frontmatter}

\title{A solution for the rare type match problem when using the DIP-STR marker system}
\vspace{0.5cm}
\author[UNIL]{G. Cereda\corref{cor1}}
\ead{giulia.cereda7@gmail.com}
\author[LEIDEN]{R. D. Gill}
\author[UNIL]{F. Taroni}
\cortext[cor1]{Corresponding author}
\address[UNIL]{University of Lausanne, School of Criminal Justice, Institute of Forensic Science, 1015 Lausanne-Dorigny, Switzerland}
\address[LEIDEN]{Leiden University, Mathematical Institute, Niels Bohrweg 1, 2333 CA Leiden, The Netherlands}

\date{\today}
\begin{abstract}

The rare type match problem is an evaluative challenging situation in which the analysis of a DNA profile reveals the presence of (at least) one allele which is not contained in the reference database. This situation is challenging because an estimate for the frequency of occurrence of the profile in a given population needs sophisticated evaluative procedures. 

The rare type match problem is very common when the DIP-STR marker system, which has proven itself very useful for dealing with unbalanced DNA mixtures, is used, essentially due to the limited size of the available database.
    The object-oriented Bayesian network proposed in \citet{cereda:2014} to assess the value of the evidence for general scenarios, was not designed to deal with this \textcolor{black}{particular} situation.
In this paper, the model is extended and partially modified to be able to calculate the full Bayesian likelihood ratio in presence of any (observed and not yet observed) allele of a given profile. The method is based on the approach developed in \citet{cereda:2015} for Y-STR data. Alternative solutions, such as the plug-in approximation and an empirical Bayesian methodology are also proposed and compared with the results obtained with the full Bayesian approach.

The paper has been published (2018) in \emph{Forensic Science International: Genetics} \textbf{34}, 88--96, \url{https://doi.org/10.1016/j.fsigen.2017.07.010}.
\end{abstract}


\end{frontmatter}

\normalem

\section{Introduction}\label{intro}
The most common task of a forensic scientist or statistician is to quantify the probative value of the observation of some scientific findings (e.g.\ DNA profiles, fragment of paint, fibres), under the hypotheses of interest for the court of justice. This is done through the quantification of the likelihood ratio.
In case the hypotheses of interest deal with whether the recovered material has the same origin as some control material, it is important to be able to quantify the rarity of the corresponding characteristics. For instance, the evidence can be the correspondence between the DNA profile of a crime stain and of a suspect: the rarer the profile, the more probative is the scientific finding regarding propositions about the source.
The rarity of the profile of interest is often used to assign the probability of the random occurrence of the given stain, and some available (and relevant) database is used to support the scientist's assignment.

The `rare type match problem', also called `the fundamental problem of forensic mathematics' \citep{brenner:2010} is the situation in which the corresponding characteristic has not been observed in the relevant reference database for the case.
One example is the DIP-STR marker system, a rather novel genotyping technique, proposed in \citet{castella:2013}, which turned out to be very useful to analyse DNA mixtures if the proportion of the DNA quantities of the, say, two contributors is more extreme than 1:10. 
Due to limited size of available databases, rare DIP-STR profiles are often encountered.

A Bayesian framework for evaluating DIP-STR results was developed in \citet{cereda:2014}, using object-oriented Bayesian networks, with the aim of calculating the likelihood ratio for mixtures of two contributors, when the major contributor's genotype is known and the two competing hypotheses are `the minor contributor is the suspect' ($h_p$) and `the minor contributor is an unknown person, unrelated to the suspect' ($h_d$), also extended to cases where the suspect is missing. 

This paper proposes a Bayesian solution for assigning the likelihood ratio for mixture results in presence of a rare type match, that is when at least one of the DIP-STR alleles of the contributors is not present in the reference database. This situation was not covered by \citet{cereda:2014}.

The Bayesian model adopted is based on a similar one proposed in~\citet{cereda:2015}.
Several issues concerning Bayesian methodology, and notation, have been improved.

The paper is structured as follows.
Section~\ref{DIPx} discusses the use of the DIP-STR marker system for extremely unbalanced mixtures, while Section~\ref{dddsa} describes the object-oriented Bayesian network that was built to evaluate DIP-STR profiling results in~\citet{cereda:2014}. 
The chosen notation and the definition of what a full Bayesian approach to likelihood ratio assessment is, can be found in Sections~\ref{notx} and~\ref{fbax}, respectively. 
The model developed to evaluate results from mixtures of two contributors in presence of the rare type match problem (described in Section~\ref{rtmps}) is detailed in Sections~\ref{sol} and~\ref{fumo}.
More detailed descriptions of the development of the full Bayesian likelihood ratio, which takes advantage of the Lemma introduced in Section~\ref{Lemmasss}, are confined to the Appendix. 
A discussion about the choice of the prior distribution for the parameters is also provided in Section~\ref{cop}, while conclusions can be found in Section~\ref{sdsdf}.

\section{DIP-STR marker system for extremely unbalanced mixtures}\label{DIPx}
A DIP-STR marker is a compound marker made of a DIP (Deletion/Insertion polymorphism, \citet[e.g.,][]{weber:2002}), and of a standard STR polymorphism. These two polymorphisms are chosen less than 500 bp apart, in order to be dependent on one another.
\textcolor{black}{It has been shown that due to very low quantity of DNA in a sample, to extreme degradation conditions, or to other physical and bio-chemical phenomena, mixtures could remain undetected using 
standard methods for the analysis of DNA mixtures \citet{oldoni:2015}}

On the other hand, as long as the minor contributor has at a specific locus at least one DIP allele different from the DIP alleles of the victim, the DIP-STR marker system allows the selected amplification of its DIP-STR genotype, up to mixture proportions as extreme as 1:1000. \textcolor{black}{Indeed, the DIP-STR markers are specifically selected to target the minor contributor, by targeting the DIP alleles not present in the victim (thus, the precise set of markers used will be case dependent). If the victim is (e.g.) homozygous L/L, then only S-alleles are targeted. Hence, by construction L-alleles of the minor will not be detected. }

At each DIP-STR locus, the possible configurations are the following (summarized in Table~\ref{Tamble1}).

\begin{itemize}
\item In the case where the major and minor contributors are DIP homozygous with different alleles (i.e., one is L-L and the other is S-S) both DIP-STR alleles of the minor contributor can be detected. This is the best scenario the scientist can be faced with.
\item If the major contributor is DIP homozygous (for instance L-L) and the minor contributor is DIP heterozygous, only one of the two DIP-STR alleles of the minor contributor can be detected: the one with the other DIP allele (in the example the allele S).
\textcolor{black}{\item If both contributors are homozygous for the same DIP alleles (both L-L or both S-S) we don't obtain results from the trace but we nevetheless obtain the information about the DIP-homozygosity of the minor.
\item The worst situation is the one in which the major is DIP heterozygous. In this case we cannot know anything about the DIP-STR profile of the minor contributor. }
\end{itemize}

\textcolor{black}{It is important to mention that, for the model presented in this paper, we assume that the DNA material is in sufficient quantity to obtain all the relevant genotypic information about the contributors that the considered set of markers is supposed to provide (i.e., no allelic drop-out, nor other artifacts.)}
\begin{table}[htbp]

\centering
\begin{tabular}{p{4.5cm} | p{5cm} |p{4.5cm}}
\hline
\textbf{DIP genotype of major/minor contributor} & \textbf{DIP-STR alleles observed in the trace, \textcolor{black}{coming from the minor contributor}}& \textbf{Information gained for the minor contributor}\\
\hline
\multirow{2}*{Hom/Hom (different kind)}    & 2 (if STR het) & \textcolor{black}{Both DIP-STR alleles}\\
   											& 1 (if STR hom)&\textcolor{black}{Both DIP-STR alleles}\\
Hom/Het 									& 1	(regardless STR)&\textcolor{black}{One of the two DIP-STR alleles}\\
Hom/Hom (same kind)						& 0 (regardless STR)&\textcolor{black}{Only DIP-homozigosity}\\
Het/Hom 									& 0 (regardless STR)&\textcolor{black}{Nothing}\\
Het/Het										& 0 (regardless STR)&\textcolor{black}{Nothing}\\
\hline
\end{tabular}
\caption{\small Informativeness of genotypic configurations. `Hom' denotes homozygous for the DIP allele, and `Het' heterozygous.}\label{Tamble1}

\end{table}

A first panel of 10 DIP-STR markers was presented in \citet{castella:2013}. A second panel with 9 additional DIP-STR markers has recently been provided in \citet{oldoni:2015}.
When one analyses a mixed stain, each of the 19 available markers may present one of the three situations described above.
\section{Bayesian network for evaluating DIP-STR profiling results from unbalanced DNA mixtures.}\label{dddsa}

\begin{figure}[htbp]

\centering
\begin{tikzpicture}

  \node [draw, ellipse]                (s1) 	at (0, -2)  { $S_1$};
  \node [draw, ellipse, minimum width=1.2cm]                (s2) 	at (2, -2)  { $S_2$};
  \node [draw, ellipse, minimum width=1.2cm]                (u1) 	at (4, -2)  { $U_1$};
  \node [draw, ellipse, minimum width=1.2cm]                (u2) 	at (6, -2)  { $U_2$};
  \node [draw, ellipse, minimum width=1.2cm]                (p1) 	at (2, -5)  { $C_1$};
  \node [draw, ellipse, minimum width=1.2cm]                (p2) 	at (4, -5)  { $C_2$};
  \node [draw, ellipse, minimum width=1.2cm]                (obs) 	at (5, -7)  {$O$};
  \node [draw, ellipse, minimum width=1.2cm]                (h) 	at (8, -3)  {$H$};
  \node [draw, ellipse, minimum width=1.2cm]                (v) 	at (8, -5)  {$V$};

  \draw[black, big arrow] (s1) -- (p1);
\draw[black, big arrow] (u1) -- (p1);
\draw[black, big arrow] (s2) -- (p2);
\draw[black, big arrow] (u2) -- (p2);
\draw[black, big arrow] (h) -- (p1);
\draw[black, big arrow] (h) -- (p2);
\draw[black, big arrow] (v) -- (obs);
\draw[black, big arrow] (p1) -- (obs);
\draw[black, big arrow] (p2) -- (obs);

 \end{tikzpicture}
\caption{\small Bayesian network corresponding to the object-oriented Bayesian network of \citet{cereda:2014}. The meaning of the nodes is described in Section~\ref{dddsa}.}\label{bnet_ch}

\end{figure}
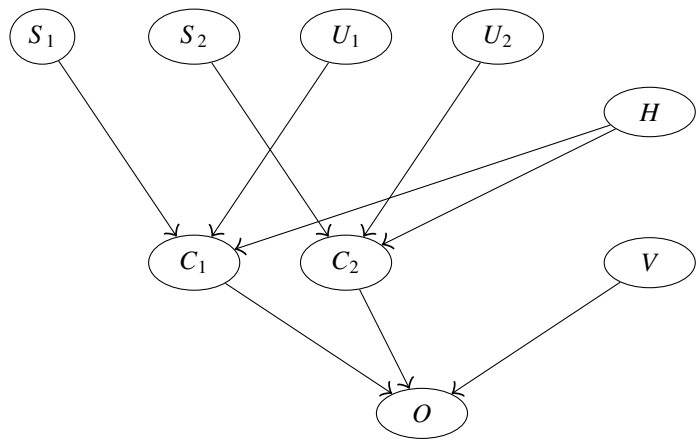


In \citet{cereda:2014} a locus specific object-oriented Bayesian network (OOBN), designed to assist the evaluation of DIP-STR results obtained from mixtures with two contributors, is proposed. The network, reproducing the mechanism described in Section~\ref{DIPx} and in Table~\ref{Tamble1}, is proposed here in the form of a Bayesian network (see Figure~\ref{bnet_ch}). It is suitable for a situation in which the DIP-STR profile of a suspect (potential contributor to the mixture) is available. The two hypotheses of interest are `the minor contributor is the suspect' ($h_p$) and `the minor contributor is an unknown person, unrelated to the suspect' ($h_d$). The major contributor is often referred to as ``the victim'', taken as a known contributor, and his/her DIP-STR profile is generally available.

It is important to notice that the only information needed from the known major contributor regards his DIP alleles. Hence, the only node in the network which concerns the victim, $V$, has three possible states: $HomoL$, $HomoS$ and $Hetero$. The node $H$ represents the two hypotheses of interest defined above.

With the exception of node $O$, the remaining part of the network deals with the unknown minor contributor. Nodes $S_1$ and $S_2$ represent the two DIP-STR alleles of the suspect. Nodes $U_1$ and $U_2$ represent the two DIP-STR alleles of the alternative (unknown) contributor in a two-person mixture. Nodes $C_1$ and $C_2$ represent the DIP-STR alleles of the actual second contributor (for example, the suspect). Depending on the state of node $H$, the second contributor's allele can be a copy of $S_1$ and $S_2$ (under state $h_p$), or of $U_1$ and $U_2$ (under state $h_d$).
The state of node $O$, which contains results obtained from the mixture, depends on the combination of $V$, $C_1$ and $C_2$ (according to Table~\ref{Tamble1}).

The probability tables for the nodes are of different types.
In the scenario considered, node $V$ is observed, since the major contributor is known. As such, its probability table is not relevant for the final result because its state is fixed, thus it is filled with equal prior probabilities for its three states. The same holds for node $H$, which is in turn instantiated to obtain the numerator and the denominator of the likelihood ratio.


Nodes $C_1$ and $C_2$ are deterministic given $H$, $S_1$, $S_2$, $U_1$, and $U_2$: if $H$ is in state $h_p$, then $C_1$ and $C_2$ are copies of, respectively, $S_1$ and $S_2$, otherwise they are copies of $U_1$ and $U_2$.
Also node $O$ is deterministic, given nodes $V$, $H$, $C_1$ and $C_2$: its probability table is filled out with 0's and 1's (according to the conditions defined in Table~\ref{Tamble1}).
The states of nodes $S_1$, $S_2$, $U_1$, $U_2$, $C_1$, and $C_2$ are $La$, $Lb$, $Lx$, $Sa$, $Sb$, $Sx$. 
Notice that at each DIP-STR locus there may be more than six possible alleles: $La$, $Lb$, $Sa$, $Sb$ are used to represent the two alleles that at most could be observed, while $Sx$ and $Lx$ represent all the other (not observed) alleles different from $a$ and $b$. 
In~\citet{cereda:2014}, this solution was preferred to having the entire list of DIP-STR alleles, in order to make the model simpler, and usable for different loci. The disadvantage is that, at each new case, the meaning of these symbols changes, and the probability tables have to be adapted accordingly. 
In this paper, we will develop a methodology to overcome this constraint.

The probability tables for nodes $S_1$, $S_2$, $U_1$, and $U_2$ should be filled with the allelic proportions corresponding to the alleles represented by names $La$, $Lb$, etc., in the population of interest. These allelic proportions are unknown, but we a have a database of DIP-STR alleles, which we can consider as a random sample from the population of interest.
In \citet{cereda:2014}, a Dirichlet distribution with all parameters equal to one was used as prior for the DIP-STR allelic proportions, and the probability tables for nodes $S_1$, $S_2$, $U_1$, $U_2$ were filled out with the posterior means (conditional to the observation of the database). However, as discussed in Section~\ref{fbax}, this approach suffers from some limitations and it can be improved and made more consistent with the Bayesian theory. Moreover, the number of possible distinct DIP-STR alleles was chosen by looking at those in the database. Thus, the model was not suitable to be used when new alleles (not previously detected) were observed. This paper aims at solving these problems. 
\section{Notation}\label{notx}

Throughout the paper the following notation is chosen: random variables and their values are denoted, respectively, with uppercase and lowercase characters: $x$ is a realization of $X$. Random vectors are denoted with bold characters: $\mathbf{x}$ is a realization of the random vector $\mathbf{X}$. Probability is denoted with $\Pr(\cdot)$, while density of a continuous random variable $X$ is denoted alternatively by $p_{X}(x)$ or by $p(x)$ when the subscript is clear from the context. For a discrete random variable $Y$, the density notation $p_Y(y)$ and the discrete one $\Pr(Y=y)$ will be alternately used.
Moreover, we will use shorthand notation like $p(y \mid x)$ to stand for the probability density of Y with respect to the conditional distribution of $Y$ given $X = x$.

Given $k\geq2$, and $\mathbf{\alpha}=(\alpha_1, ..., \alpha_k)$ such that $\alpha_i>0$,
$$\mathbf{X}\sim \text{Dir}^k(\alpha_1, ..., \alpha_{k})$$ means that vector $\mathbf{x}$ follows a $k$-dimensional Dirichlet distribution \citep{press:2009}, whose density is
$$p(\mathbf{x})=\frac{\Gamma(\sum_{i=1}^k \alpha_i)}{\prod_{i=1}^k\Gamma(\alpha_i)}\prod_{i=1}^kx_i^{\alpha_i-1}.$$

In the appendix, we will denote with $\mathbf{z}=(\mathbf{x}, y)$ the vector $\mathbf{z}$ obtained by adding element $y$ at the end of vector $\mathbf{x}$.

\section{Full Bayesian approach}~\label{fbax}
In the case of interest, for each analysed locus the forensic scientist or statistician is given the following input data: the victim's and the suspect's DIP-STR profile (denoted as $E_v$ and $E_s$ respectively), along with the DIP-STR alleles obtained from the mixture ($E_m$). This data has to be evaluated in the light of the hypotheses of interest ($h_p$ and $h_d$) as defined in Section~\ref{intro}.
The evaluation of such evidence heavily depends on the allelic proportions of the DIP-STR alleles of the trace and of the suspect, which are unknown. The vector $\boldsymbol{\theta}$, containing the population proportions of all the possible DIP-STR alleles at the considered locus, is the nuisance parameter of the model. 
A database (denoted here as $D$), consisting of a list of DIP-STR alleles from the population of interest is given to the statistician, in order to support him in the assessment of the uncertainty about $\boldsymbol{\theta}$. 
The data to evaluate are thus made of $E$=($E_v$, $E_s$, $E_m$) and $D$. This notation reflects the distinction described in \citet{cereda:2015} between `evidence', data directly related to the crime, and `background', data related only to the nuisance parameter of the model. 

The full Bayesian approach consists of modelling all these variables, including $\boldsymbol{\theta}$, as random variables whose joint distribution $\Pr$ reflects prior belief of the expert. 

The largely accepted method to evaluate the data in order to discriminate between the two hypotheses of interest, is the calculation of the \emph{Bayes factor} (BF), in forensic context regularly called \emph{likelihood ratio} (LR). It is defined as the ratio of the probabilities of observing the data under the two competing hypotheses:
\begin{equation}
\label{LR}
\LR=\frac{\Pr(E=e, D=d \mid H= h_{p})}{\Pr(E=e,D=d \mid H=h_{d})}=\frac{\Pr(E=e \mid D=d, H=h_{p})}{\Pr(E=e\mid D=d, H=h_{d})},
\end{equation}
where the last equality holds in virtue of the independence of database and hypotheses.

The nuisance parameter $\boldsymbol{\theta}$ has been integrated out according to its prior distribution. Notice indeed that $\boldsymbol{\theta}$ does not appear in \eqref{LR}.

In \citet{cereda:2014}, we used a different approach: Bayesian estimates of the allelic proportions were plugged into the probability tables for nodes $S_1$, $S_2$, $U_1$, and $U_2$. This is equivalent to using a likelihood ratio for a given $\boldsymbol{\theta}$, such as 
$$
\LR=\frac{\Pr(E=e\mid \boldsymbol{\Theta} = \boldsymbol{\theta}, D=d, H=h_p)}{\Pr(E=e\mid \boldsymbol{\Theta}=\boldsymbol{\theta}, D=d, H=h_d)},$$
and to plug inside the estimates for $\boldsymbol{\theta}$.

The plug-in method can be seen as an approximation to the full Bayesian method \citep{cereda:2015}. To obtain it, a Bayesian network is built which allows one to use an integrated full Bayesian approach, by introducing, among others, a node which represents the database $D$, and a node that represents the nuisance parameter $\boldsymbol{\theta}$.  
The full Bayesian approach is then compared to the plug-in method, to check the impact of the approximations.

\section{Rare type match problem}\label{rtmps}

When the findings to evaluate include a correspondence between the DNA profile of a particular piece of evidence (i.e., a trace of unknown origin) and a suspect's DNA profile, but at least one of the alleles of this profile are not present in the available database, it is difficult to assess the uncertainty over the population proportion of that allele. It is likely to be a rare allele (from which the term \emph{rare type match problem}) but it is challenging to quantify how rare. This assessment is important for the quantification of the likelihood ratio: the rarer the matching profile, the larger is the likelihood ratio. 

Using DIP-STR data, it is very likely to encounter the rare type match problem, because the available database size is still limited \citep{oldoni:2015}.
The same happens when Y-chromosome (or mitochondrial) DNA profiles are used, since the set of possible Y-STR profiles is extremely
large. As a consequence, most of the Y-STR haplotypes are not represented in the database.
In \citet{cereda:2015, cereda:2015b, cereda:2015c} several (Bayesian and frequentist) solutions are proposed for the rare type match problem for Y-STR data. 
The object-oriented Bayesian network of~\citet{cereda:2014}, here presented in Figure~\ref{bnet_ch}, cannot be used in the case of a rare type match problem: there, the number of different alleles at a given locus was considered as fixed, and equal to that observed in the database. This makes that model useless in cases where new DIP-STR alleles are observed.

As a solution, we will consider the number of different DIP-STR alleles present in the population as random, by introducing additional variables in the model, explained in detail in Section~\ref{sol}. This is based on one of the Bayesian methods proposed in~\citet{cereda:2015}.

\section{A prior for $\boldsymbol{\theta}$}\label{sol}

\textcolor{black}{Let us denote with L-STR (or S-STR) the DIP-STR alleles which have the DIP part equal to L (or S).}
Assume that, at a specific locus, there are at most $m$ theoretically possible L-STR alleles and $m$ theoretically possible S-STR alleles. 
The random vector $\boldsymbol{\Theta}=(\Theta^L_{1}, ..., \Theta^L_{m}, \Theta^S_{1}, ..., \Theta^S_{m} )$ contains the population proportions of all the potential $2m$ DIP-STR alleles at that locus (for instance, alphabetically ordered).

Only $k^L$ ($k^S$) of the $m$ possible L-STR (S-STR) alleles are actually present in nature (or more specifically in the population of interest), but $k^{L}$ and $k^{S}$ are unknown. 
Which of the $m$ L-STR alleles are those $k^{L}$ and $k^{S}$ is not known either.
 
The vector $\mathbf{t}^L$ contains the ordered positions (from 1 to $m$) of the $k^{L}$ L-STR alleles present in the population of interest. $\mathbf{t}^L$ is modelled through a random variable $\mathbf{T}^L$: each possible configuration $\mathbf{t}^L$ is assumed as equiprobable, hence it is chosen uniformly at random from the possible $\binom{m}{k^{L}}$ configurations. 
Random vector $\mathbf{T}^S$ is defined similarly.
Notice that $\theta^L_i=0, \forall i \notin \mathbf{t}^L$, and $\theta^S_i=0, \forall i \notin \mathbf{t}^S$.

\begin{figure}[htbp]

\centering
\begin{tikzpicture}
  \node [draw, ellipse,  minimum width=1.2cm]                (kl) at (-1, 0.5)  {$k^L$};
  \node [draw, ellipse, minimum width=1.2cm]                (ks) at (1, 0.5)  {$k^S$}; 
  \node [draw, ellipse, minimum width=1.2cm]                (tl) at (-1, -1.5)  {$\mathbf{t}^L$}; 
  \node [draw, ellipse, minimum width=1.2cm]                (ts) at (1, -1.5)  {$\mathbf{t}^S$}; 
  \node [draw, ellipse, minimum width=1.2cm]                (phil) at (-1, -3.5)  {$\boldsymbol{\Phi}^L$}; 
  \node [draw, ellipse, minimum width=1.2cm]                (phis) at (1, -3.5)  {$\boldsymbol{\Phi}^S$}; 
  \node [draw, ellipse, minimum width=1.2cm]                (theta) at (0, -5.5)  {$\boldsymbol{\Theta}$}; 
 \node [draw, ellipse, minimum width=1.2cm]                (psi) at (3, -3.5)  {$\Psi$}; 
\draw[black, big arrow] (kl) -- (tl);
\draw[black, big arrow] (ks) -- (ts);
\draw[black, big arrow] (tl) -- (phil);
\draw[black, big arrow] (ts) -- (phis);

\draw[black, big arrow] (phil) -- (theta);
\draw[black, big arrow] (phis) -- (theta);
\draw[black, big arrow] (psi) -- (theta);
 \end{tikzpicture}
\caption{\small The conditional dependency relation of the random variables used to build the distribution of $\boldsymbol{\theta}$. The definition of the nodes can be found in Section~\ref{sol}.}\label{bnet_ch2vc}

\end{figure}
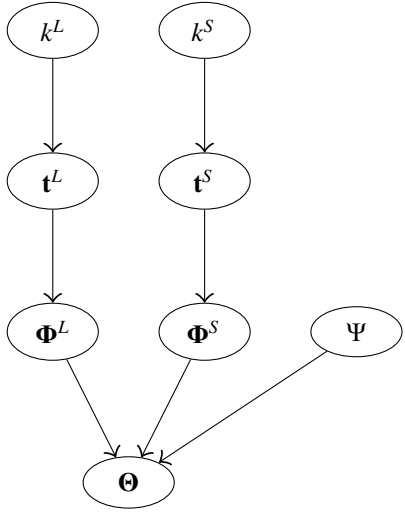

Specifying $\boldsymbol{\Theta}$ is equivalent to specifying three random variables $\boldsymbol{\Phi}^L$, $\boldsymbol{\Phi}^S$, $\Psi$. $\Psi$ is the sum of the occurrence probabilities of the L-STR alleles
$$ \psi=\sum_{i=1}^m \theta^L_i, $$
while $\boldsymbol{\phi}^L$ is the normalized vector of the occurrence probabilities of the L-STR alleles. Stated otherwise, 
$$
\boldsymbol{\phi}^L=(\frac{\theta^L_1}{\psi}, ..., \frac{\theta^L_m}{\psi}). 
$$
Similarly, $\boldsymbol{\phi}^S$ is the normalized vector of the relative frequencies of the S-STR alleles:
$$
\boldsymbol{\phi}^S=(\frac{\theta^S_{1}}{1 - \psi}, ..., \frac{\theta^S_{m}}{1 - \psi}). 
$$
The prior distribution for $\boldsymbol{\theta}$ can be described in terms of the prior over $\boldsymbol{\Phi}^L$, $\boldsymbol{\Phi}^S$, and $\Psi$, which will be taken to be independent. The latter is distributed according to a Beta(1,1), while the positive entries of $\boldsymbol{\phi}^L$, i.e., $(\phi_i^L \mid i \in \mathbf{t}^L)$ are Dirichlet distributed, given $\mathbf{t}^L$, with all hyperparameters equal to $\alpha$. The same holds for $\boldsymbol{\phi}^S$ given $\mathbf{t}^S$. Hence, the distribution of node $\boldsymbol{\theta}$ can be described in terms of the distribution of seven additional random variables, whose conditional dependencies can be described by the Bayesian network of Figure~\ref{bnet_ch2vc}.
Bayesian networks using beta and Dirichlet distributions in forensic contexts are presented in \citet{biedermann:2011b}. Other examples can also be found in \citet{taroni:2014}.

\section{Full model}\label{fumo}

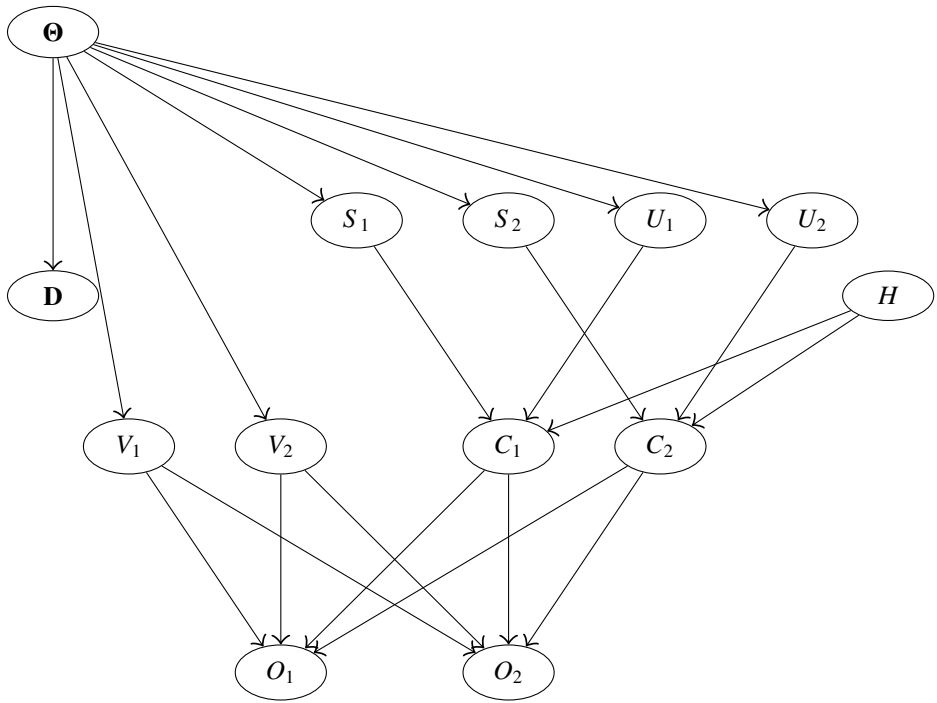
\begin{figure}[htbp]

\centering
\begin{tikzpicture}
  \node [draw, ellipse, minimum width=1.2cm]                (theta) at (-2, 0.5)  {$ \boldsymbol{\Theta}$};
  \node [draw, ellipse, minimum width=1.2cm]                (b) 	at (-2, -3) { $\mathbf{D}$};
  \node [draw, ellipse, minimum width=1.2cm]                (s1) 	at (2, -2)  { $S_1$};
  \node [draw, ellipse, minimum width=1.2cm]                (s2) 	at (4, -2)  { $S_2$};
  \node [draw, ellipse, minimum width=1.2cm]                (u1) 	at (6, -2)  { $U_1$};
  \node [draw, ellipse, minimum width=1.2cm]                (u2) 	at (8, -2)  { $U_2$};
  \node [draw, ellipse, minimum width=1.2cm]                (p1) 	at (4, -5)  { $C_1$};
  \node [draw, ellipse, minimum width=1.2cm]                (p2) 	at (6, -5)  { $C_2$};
  \node [draw, ellipse, minimum width=1.2cm]                (obs) 	at (1, -8)  {$O_1$};
    \node [draw, ellipse, minimum width=1.2cm]               (obs2) 	at (4, -8)  {$O_2$};
  \node [draw, ellipse, minimum width=1.2cm]                (h) 	at (9, -3)  {$H$};
  \node [draw, ellipse, minimum width=1.2cm]                (v) 	at (-1, -5)  {$V_1$};
  \node [draw, ellipse, minimum width=1.2cm]                (v2) 	at (1, -5)  {$V_2$};



  \draw[black, big arrow] (theta) -- (s1);  
  \draw[black, big arrow] (theta) -- (s2);
  \draw[black, big arrow] (theta) -- (u1);
  \draw[black, big arrow] (theta) -- (u2);

  \draw[black, big arrow] (theta) -- (v);
  \draw[black, big arrow] (theta) -- (v2);
  \draw[black, big arrow] (theta) -- (b);
  \draw[black, big arrow] (s1) -- (p1);
\draw[black, big arrow] (u1) -- (p1);
\draw[black, big arrow] (s2) -- (p2);
\draw[black, big arrow] (u2) -- (p2);
\draw[black, big arrow] (h) -- (p1);
\draw[black, big arrow] (h) -- (p2);
\draw[black, big arrow] (v) -- (obs);
\draw[black, big arrow] (v) -- (obs2);
\draw[black, big arrow] (v2) -- (obs);
\draw[black, big arrow] (v2) -- (obs2);
\draw[black, big arrow] (p1) -- (obs);
\draw[black, big arrow] (p2) -- (obs);
\draw[black, big arrow] (p1) -- (obs2);
\draw[black, big arrow] (p2) -- (obs2);

 \end{tikzpicture}
\caption{\small Bayesian network for the Dirichlet-multinomial model with a random number of types, to be used for DIP-STR data. The definition of the nodes can be found in Section~\ref{fumo}.}\label{bnet_ch2}

\end{figure}

This model is represented by the Bayesian network of Figure~\ref{bnet_ch2}. Notice that there are differences from the model depicted in Figure~\ref{bnet_ch}, among which is the presence of node $\boldsymbol{\theta}$ distributed as described in Section~\ref{sol}. The first difference lies in the definition of nodes $S_1$, $S_2$, $U_1$, $U_2$, $C_1$, and $C_2$. Their values are couples (L,$i$) or (S,$i$) where $i\in \{1, ..., m\}$, describing the position in $\boldsymbol{\theta}$ of the corresponding DIP-STR allele. The same holds for nodes $V_1$ and $V_2$, which replace node $V$ of Figure~\ref{bnet_ch}, and represent the two DIP-STR alleles of the victim. All these nodes are now linked to node $\boldsymbol{\Theta}$ because, 
given $\boldsymbol{\Theta}=\boldsymbol{\theta}$, the random variables $S_1$, $S_2$, $U_1$, $U_2$, $V_1$, $V_2$ have the following density (with parameter $\boldsymbol{\theta}$):
\begin{align} \label{fed}
p( (j,i) \mid \boldsymbol{\theta})= \theta^j_i, \quad \forall j\in\{L,S\}, \forall i\in \{1,..., m\}.
\end{align}
The second difference is the presence of two nodes $O_1$ and $O_2$, instead of a single node $O$ as in Figure~\ref{bnet_ch}. $O_1$ represents one of the DIP-STR alleles observed from the mixture (if any, 0 otherwise). $O_2$ is always 0 unless we are in the situation described by the first row of Table~\ref{Tamble1}, where two DIP-STR alleles are observed. 
In this case, the convention is for $O_1$ and $O_2$ to be ordered alphabetically.

The random vector $\mathbf{D}$ represents the available database of size $n$, through the list of labels ($(L,i)$ or $(S,i)$) of the DIP-STR alleles contained in the database. The order does not matter, so we can choose the order in which the alleles appear in the database. A particular configuration of $\mathbf{D}$ is denoted as $\mathbf{d}= (d_1, ..., d_{n})$, where, given $\boldsymbol{\theta}$, each component is i.i.d.\ with the same density as in \eqref{fed}.

According to this notation, the likelihood ratio for the scenario of interest can be written as 
\begin{equation}\label{eqws}
\LR =\frac{p(o_1, o_2, s_1, s_2, v_1, v_2, \mathbf{d}\mid h_p)}{p(o_1, o_2, s_1, s_2, v_1, v_2, \mathbf{d}\mid h_d)}=\frac{p(o_1, o_2 \mid s_1, s_2, v_1, v_2, \mathbf{d}, h_p)}{p(o_1, o_2 \mid s_1, s_2, v_1, v_2, \mathbf{d}, h_d)}. 
\end{equation}
Due to the complexity of the chosen distribution, the Bayesian network cannot be treated with available software, such as Hugin, or OpenBUGS. However, the likelihood ratio can be obtained analytically using the Lemma presented in Section~\ref{Lemmasss}.

\section{Lemma}\label{Lemmasss}

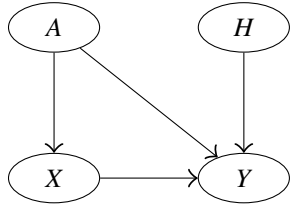
\begin{figure}[htbp]
\centering
  \begin{tikzpicture}
\node[draw, ellipse, minimum width=1.2cm]                (t) at (-1,0)  { $A$};
\node [draw, ellipse, minimum width=1.2cm]              (h) at (1.5,0) {$H$};
 \node [draw, ellipse, minimum width=1.2cm]              (d) at (-1,-2) { $X$};
\node [draw, ellipse, minimum width=1.2cm]              (dr) at (1.5,-2) { $Y$};
  \draw[black, big arrow]  (t) -- (d);
  \draw[black, big arrow] (h) -- (dr);     
  \draw[black, big arrow]  (t) -- (dr);
   \draw[black, big arrow] (d) -- (dr);
\end{tikzpicture}

     \caption{\small Conditional dependencies of the random variables of the Lemma}\label{figure1}

     \end{figure}

   \begin{lemma}\label{Lemma1}

Given four random variables $A$, $H$, $X$ and $Y$, whose conditional dependencies are represented by the Bayesian network of Figure~\ref{figure1}, the likelihood function for $h$, given $X=x$ and $Y=y$ satisfies 
$$\mathrm{lik}(h\mid x, y)~ \propto \mathbb{E}(p(y \mid x, A, h) \mid X = x).$$   \end{lemma}  

This Lemma, proven in \citet{cereda:2015c}, is very general: it applies to every group of random variables whose conditional dependencies are represented by the Bayesian network of Figure~\ref{figure1}, and it is very useful due to the possibility of applying it to a very common forensic situation: the prosecution and the defence disagree on the distribution of part of the data ($Y$) but agree on the distribution of the other part ($X$), when the distribution of $X$ and $Y$ depends on some parameters ($A$).
This Lemma can also be used for the DIP-STR model presented in Section~\ref{sol}. However, it is not straightforward to identify in the Bayesian network of Figure~\ref{bnet_ch2} the required structure shown in Figure~\ref{figure1}.
Luckily, the same model can be represented in several ways: we will propose a modification of the Bayesian network of Figure~\ref{bnet_ch2} into something which more clearly shows the required structure. This will be done in two steps: first, we will remove unnecessary nodes, and then we will group some of the others. 
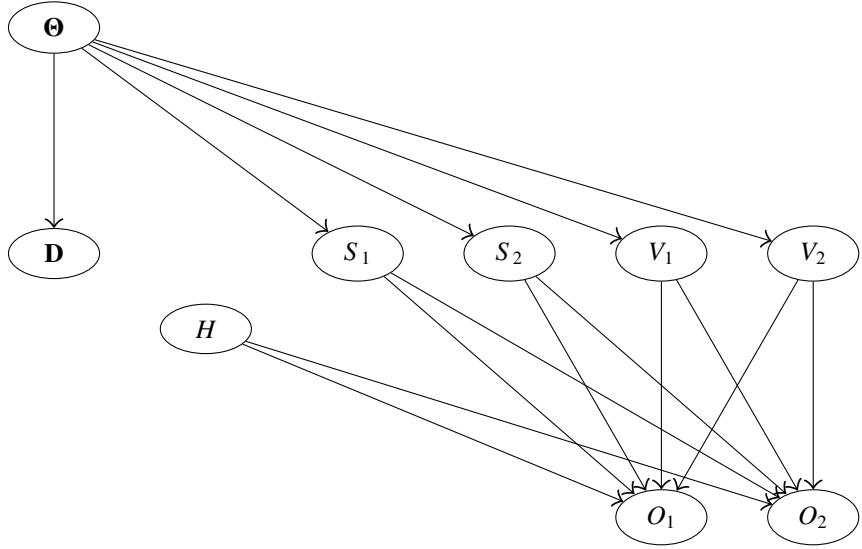
\begin{figure}[htbp]

\centering
\begin{tikzpicture}
  \node [draw, ellipse, minimum width=1.2cm]                (theta) at (-2, 1)  {$ \boldsymbol{\Theta}$};
  \node [draw, ellipse, minimum width=1.2cm]                (b) 	at (-2, -2) { $\mathbf{D}$};
    \node [draw, ellipse, minimum width=1.2cm]                (v1) 	at (6, -2)  { $V_1$};
  \node [draw, ellipse, minimum width=1.2cm]                (v2) 	at (8, -2)  { $V_2$};
  \node [draw, ellipse, minimum width=1.2cm]                (s1) 	at (2, -2)  { $S_1$};
  \node [draw, ellipse, minimum width=1.2cm]                (s2) 	at (4, -2)  { $S_2$};

  \node [draw, ellipse, minimum width=1.2cm]                (obs) 	at (6, -5.5)  {$O_1$};
    \node [draw, ellipse, minimum width=1.2cm]               (obs2) 	at (8, -5.5)  {$O_2$};
  \node [draw, ellipse, minimum width=1.2cm]                (h) 	at (0, -3)  {$H$};



  \draw[black, big arrow] (theta) -- (s1);  
  \draw[black, big arrow] (theta) -- (s2);

  \draw[black, big arrow] (theta) -- (b);
   \draw[black, big arrow] (theta) -- (v1); 
   \draw[black, big arrow] (theta) -- (v2);
  \draw[black, big arrow] (s1) -- (obs);
  \draw[black, big arrow] (s1) -- (obs2);
\draw[black, big arrow] (s2) -- (obs2);
\draw[black, big arrow] (s2) -- (obs);
\draw[black, big arrow] (h) -- (obs);
\draw[black, big arrow] (h) -- (obs2);

\draw[black, big arrow] (v1) -- (obs);
\draw[black, big arrow] (v2) -- (obs);

\draw[black, big arrow] (v1) -- (obs2);
\draw[black, big arrow] (v2) -- (obs2);

 \end{tikzpicture}
\caption{\small An alternative representation of the DIP-STR mixture model presented in Figure~\ref{bnet_ch2}.}\label{bnet_ch22}

\end{figure}

\paragraph{Step 1.} The Bayesian network presented in Figure~\ref{bnet_ch22} is obtained by removing from the Bayesian network of Figure~\ref{bnet_ch2} nodes $U_1$, $U_2$, $C_1$, and $C_2$. The conditional probability tables of nodes $O_1$ and $O_2$ can be directly expressed in terms of $S_1$, $S_2$, $V_1$, $V_2$, and $H$, in a way that makes the model of Figure~\ref{bnet_ch22} equivalent to the previous one (Figure~\ref{bnet_ch2}). 

\begin{figure}[htbp]

\centering
\begin{tikzpicture}
  \node [draw, ellipse, minimum width=1.2cm]                (theta) at (-2, 1)  {$ \boldsymbol{\Theta}$};
  \node [draw, ellipse, minimum width=1.2cm]                (b) 	at (-2, -1) { $\mathbf{B}$};

  \node [draw, ellipse, minimum width=1.2cm]                (obs) 	at (0, -2.5)  {$O_1$};
    \node [draw, ellipse, minimum width=1.2cm]               (obs2) 	at (2, -2.5)  {$O_2$};
  \node [draw, ellipse, minimum width=1.2cm]                (h) 	at (4, 0)  {$H$};
 \node [ minimum size=0.5cm]		(x) 	at (-2.9, 1.4) { $A$};
\node [ minimum size=0.5cm]		(x) 	at (3, -2.1) { $Y$};
\node [ minimum size=0.5cm]		(x) 	at (-2.9, -0.6) { $X$};

             \draw [red, dashed] (-1.4,-0.4) -- (-1.4,-1.6);
                \draw [red, dashed] (-2.6,-1.6) -- (-1.4,-1.6);

 \draw [red, dashed] (-2.6,-0.4) -- (-2.6,-1.6);
 \draw [red, dashed] (-2.6,-0.4) -- (-1.4,-0.4);

\draw [red, dashed] (-2.6, 1.6) -- (-1.4, 1.6);
\draw [red, dashed] (-2.6, 1.6) -- (-2.6,0.3);
\draw [red, dashed] (-2.6, 0.3) -- (-1.4,0.3);
\draw [red, dashed] (-1.4, 1.6) -- (-1.4,0.3);

\draw [red, dashed] (-0.7, -1.9) -- (2.7, -1.9);
\draw [red, dashed] (-0.7, -3.1) -- (-0.7, -1.9);
\draw [red, dashed] (-0.7, -3.1) -- (2.7, -3.1);
\draw [red, dashed] (2.7, -3.1) -- (2.7, -1.9);


  \draw[black, big arrow] (theta) -- (b);
\draw[black, big arrow] (b) -- (obs);
\draw[black, big arrow] (b) -- (obs2);

  \draw[black, big arrow] (theta) -- (obs);
  \draw[black, big arrow] (theta) -- (obs2);
\draw[black, big arrow] (h) -- (obs);
\draw[black, big arrow] (h) -- (obs2);

 \end{tikzpicture}
\caption{\small A simpler structure for the Bayesian network, suitable to be used for the Lemma. Dashed lines show the choice for the corresponding variables $A$, $X$, and $Y$ of Figure~\ref{figure1}.}\label{bnet_ch221}

\end{figure}
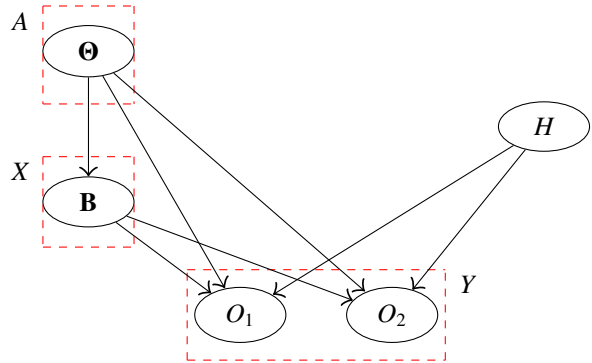

\paragraph{Step 2.} The Bayesian network of Figure~\ref{bnet_ch221} can be obtained by substituting some of the nodes of the Bayesian network of Figure~\ref{bnet_ch22} with a single node. Indeed, instead of having the random vector $\mathbf{D}$ and four additional random variables ($S_1$, $S_2$, $V_1$, and $V_2$), we can group all these together into a random vector $\mathbf{B}$, of length $n + 4$. The first $n$ elements are the labels contained in $\mathbf{D}$, the fourth to last and third to last are the labels in $S_1$ and $S_2$, while the second to last and the last are the labels in $V_1$, and $V_2$, respectively.


The Bayesian network of Figure~\ref{bnet_ch221} can be used to represent the same model as that represented by Figure~\ref{bnet_ch}, by carefully adapting the conditional distribution of $O_1$, and $O_2$. We can apply the Lemma to our model by defining $Y=(O_1, O_2)$, $X=\mathbf{B}$, and $A=\Theta$. This leads to
$$\LR=\frac{p(o_1, o_2, \mathbf{b}\mid h_p)}{p(o_1, o_2, \mathbf{b}\mid h_d)}=\frac{\textrm{lik}(h_p\mid o_1, o_2,\mathbf{b})}{\textrm{lik}(h_d\mid o_1, o_2,\mathbf{b})}=\frac{\mathbb{E}(p(o_1, o_2 \mid \mathbf{b}, \boldsymbol{\Theta}, h_p) \mid \mathbf{B} = \mathbf{b})}{\mathbb{E}(p(o_1, o_2 \mid \mathbf{b}, \boldsymbol{\Theta}, h_d) \mid \mathbf{B} = \mathbf{b})}.$$

Notice that we assume that under the prosecution's hypothesis, $p(o_1, o_2 \mid \mathbf{b}, \boldsymbol{\Theta}, h_p)=1$.
Therefore, the likelihood ratio can be simplified:
\begin{equation}\label{eq1e}
\LR=\frac{1}{\mathbb{E}(p(o_1, o_2 \mid \mathbf{b}, \boldsymbol{\Theta}, h_d) \mid \mathbf{B} = \mathbf{b})}.
\end{equation}

\textcolor{black}{Indeed, as one would expect, under Hd the probability of seeing $o_1$ and $o_2$ is given by the expected frequency of that combination, given the alleles in the database, of the suspect and of the victim which have been observed. }

\begin{table}[htbp]

\centering
\begin{tabular}{|p{4cm}  | p{1cm} | p{1cm} | p{3.5cm}|}
\hline
\textbf{Victim's DIP alleles} &  $o_1$ & $o_2$ &$p(o_1, o_2 \mid \mathbf{b}, \boldsymbol{\Theta}, h_d)$\\[4.5pt]
\hline
\multirow{3}*{L-L}    	
&   (S, $i$)  & (S, $j$) & $2\Theta^S_{i}\Theta^S_j$\\[4.5pt]
&    (S, $i$) & 0 & $(\Theta^{S}_{i})^2+2 \Theta^S_i \Psi$\\[4.5pt]
&   0 & 0 & $\Psi^2$\\[4.5pt]
\hline
\multirow{3}*{S-S}    	&  (L, $i$) & (L, $j$) & $\textcolor{black}{2}\Theta^L_{i}\Theta^L_j$\\[4.5pt]
				&  (L, $i$) & 0 & $(\Theta^L_{i})^2+2\Theta^L_i(1-\Psi)$\\[4.5pt]
				 &  0 & 0 & $(1-\Psi)^2$\\[4.5pt]

\hline
\end{tabular}
\caption{\small Different forms that $p(o_1, o_2 \mid \mathbf{b}, \boldsymbol{\Theta}, h_d)$ can take, based on the DIP-STR alleles observed from the trace and on the victim's DIP alleles.  The case in which the victim is heterozygous is not of interest.} \label{Table2ed}

\end{table}
$p(o_1, o_2 \mid \mathbf{b}, \boldsymbol{\Theta}, h_d)$ is a function of some components of the vector $\boldsymbol{\Theta}$. The form of this function depends on the combination of the DIP-STR alleles of the victim and of the trace (see Table~\ref{Table2ed}).
The expectation in the denominator of \eqref{eq1e} is to be taken using the posterior distribution $\boldsymbol{\Theta} \mid \mathbf{B}=\mathbf{b}$. \textcolor{black}{This is developed in detail in the Appendix, leading to the following 5 relevant equations:}

\begin{equation*}
\mathbb{E}(\Psi^2\mid \mathbf{b} )= \frac{(n^{L}+1)(n^{L}+2)}{(n+6)(n+7)},
\end{equation*}

\begin{equation*}
\mathbb{E}((1-\Psi)^2\mid \mathbf{b} )=\frac{(n^{S}+1)(n^{S}+2)}{(n+6)(n+7)}.
\end{equation*}

\begin{equation*}
\mathbb{E}(\Theta^L_i\Theta^L_j \mid \mathbf{b} )=(\alpha+{n^L_{i}})(\alpha+{n^{L}_j}) \frac{\sum_{k= k^L_{\mathbf{b}}}^{m} w^L(k) g^L(k) }{\sum_{k= k^L_{\mathbf{b}}}^{m} w^L(k)}\frac{(n^{L}+1)(n^{L}+2)}{(n+6)(n+7)},  \end{equation*}

\begin{equation*}
\mathbb{E}((\Theta_i^L)^2 \mid \mathbf{b} )=
(\alpha+n_{i}^L)(\alpha+n_{i}^L+1)\frac{\sum_{k= k^L_{\mathbf{b}}}^{m} w^L(k) g^L(k) }{\sum_{k= k^L_{\mathbf{b}}}^{m} w^L(k)}\frac{(n^{L}+1)(n^{L}+2)}{(n+6)(n+7)}, \end{equation*}

\begin{equation*}
\mathbb{E}(\Theta_i^L(1-\Psi) \mid \mathbf{b} )=
(\alpha+n^L_{i})\frac{\sum_{k= k^L_{\mathbf{b}}}^{m} \frac{w^L(k)}{k\alpha + n^L} }{\sum_{k= k^L_{\mathbf{b}}}^{m} w^L(k)}\frac{(n^L+1)(n^S+1)}{(n+6)(n+7)},  \end{equation*}
where $w^L(k)=\binom{k}{k^L_{\mathbf{b}}}p(k) \frac{\Gamma(k\alpha)}{\Gamma(n^L+k\alpha)}$, and $g^L(k)=\frac{1}{(k\alpha+n^L)(k\alpha+n^L+1)}$. The meaning of $n$,$n^{L}$, $n^{S}$, and $k^L_{\mathbf{b}}$ can be found in Table \ref{Table4w44}.

where $w^L(k)=\binom{k}{k^L_{\mathbf{b}}}p(k) \frac{\Gamma(k\alpha)}{\Gamma(n^L+k\alpha)}$, and $g^L(k)=\frac{1}{(k\alpha+n^L)(k\alpha+n^L+1)}$. The meaning of $n$,$n^{L}$, $n^{S}$, and $k^L_{\mathbf{b}}$ can be found in Table \ref{Table4w44}.

\section{Choice of priors}\label{cop}
The Appendix shows the form of the denominator of the likelihood ratio for the different cases which one may encounter (for any $m$, any parameter $\alpha>0$ for the Dirichlet distribution, and any prior $p(k)$ over $k^L$ and $k^S$). The choice of a value for $\alpha$, $m$, and of a prior over $k^L$ is very delicate. If the expert has strong opinions about the number of L-STR (S-STR) alleles potentially present in nature ($m$) and in the population of interest ($k^L$ and $k^S$), he can choose a prior which reflects his beliefs. Otherwise, he can try to use classical priors such as the Poisson distribution, the Negative binomial distribution (both of them truncated so as to have support only over $\{1,..., m\}$), or the uniform prior over $\{1,..., m\}$. 


\subsection{Alternative solutions}
The most natural choice is to give a uniform prior (over $\{1, ..., m\}$) to $k^L$ and $k^S$, combined with that of having  all the $k^{L}+k^{S}$ hyperparameters of the Dirichlet priors over $\boldsymbol{\phi}_\mathbf{b}^L$ and $\boldsymbol{\phi}_\mathbf{b}^S$ equal one another. These choices represent the lack of knowledge on the number of categories and make the computations tractable.

One of the limitations of having all the hyperparameters $\alpha$ equal one another is that the posterior for $k^L$, given $\mathbf{b}$ uses only the number of distinct alleles of type L as information, and ignores other useful information contained in $\mathbf{b}$.
An alternative solution, which compensates for this undesired feature, consists of estimating $k^L$ through the database, instead of putting a prior on it. This can be called an empirical Bayesian approach. Notice that such an undesired situation does not appear if personal beliefs are used to specify the prior distribution. 
Let us define the vector $\boldsymbol{\phi_{\mathbf{b}}}^L$ made of the allelic proportions of the L-STR alleles observed in the augmented database, and of a last component $\bar{\phi}^L_{\mathbf{b}}$ which is the sum of the allelic proportions of all the L-STR alleles not observed in $\mathbf{b}$. \textcolor{black}{$\bar{\phi}^L_{\mathbf{b}}$ is the probability of observing a new L-STR allele in the $n+1$th draw from the population}. $\boldsymbol{\phi_{\mathbf{b}}}^L$ given $k^L$ and \textcolor{black}{$\mathbf{b}$} is Dirichlet distributed, hence we can obtain the posterior expected values of $\bar{\phi}^L_{\mathbf{b}}$:
\begin{equation}\label{eqwsd}
\mathbb{E}(\bar{\phi}^L_{\mathbf{b}}\mid k^L, \mathbf{b}) = \frac{(k^L-k_{\mathbf{b}}^L)\alpha}{k^L\alpha + n^L}.
\end{equation}

The so-called \emph{Good-Turing estimator} \citep{good:1953} says that the expected value for the probability of the unobserved types can be approximated by the proportion of L-STR singletons (i.e, alleles observed only once) in the database. Stated otherwise, 
\begin{equation}\label{eqwsd2}
\mathbb{E}(\bar{\phi}^L_{\mathbf{b}}|k^L, \mathbf{b} ) \approx \frac{n_1^L}{n^L}.
\end{equation}
where $n_1^L$ is the number of DIP-STR alleles observed only once in the augmented database. The two quantities \eqref{eqwsd} and \eqref{eqwsd2} can be equated in order to obtain an empirical Bayesian estimate of $k^L$ as 
$$\hat{k^L}=\frac{n_1^L n^L+k_{\mathbf{b}}^L\alpha n^L}{\alpha n^L-\alpha n_1^L}.$$
The likelihood ratio for this choice can be obtained using the same formulas developed in the Appendix by using prior over $k^L$ the degenerate prior which gives a probability of one to value $\widehat{k^L}$. This solution allows one to use more information ($n_1^L$ and $n_1^S$) from $\mathbf{b}$.

The third option is to use the plug-in approximation proposed by some literature, which estimates the allelic frequencies by their posterior expectation, after the observation of a database. 
One of the aims of this paper is to investigate the goodness of this approximation, in the case of a rare type match problem. 

%

We did some experiments using marker MID1950-D20S473 \citep{castella:2013}, and considering the two cases described in Table~\ref{acsesd}.

%
%
%
%
%

\begin{table}[htbp]

\centering
\begin{tabular}{|c|c|c|c|c|c|c|c|}
\hline
		&Victim's alleles	& Suspect's alleles	& Observed alleles	&$n^L$& $n_1^L$ &$k_{\mathbf{b}}^L$ 	\\
		\hline
Case 1 	&  S11 - S11		& L2 - L13  		& L2 - L13		& 81 & 2 & 6 	\\
Case 2 	&  S11 - S11		& L2 - S12   		& L2 			& 80 & 2 & 6	\\
\hline
\end{tabular}

\caption{\small Allelic configurations of the victim and of the suspect in the two cases of interest, at marker MID1950-D20S473.}

\label{acsesd}

\end{table}%

Allele L2 was not contained in the database of reference, hence we are in presence of the rare type match case. The available database, augmented with the victim's and the suspect's DIP-STR alleles, contains 11 different DIP-STR alleles, for a total number of 210 observations (from 105 individuals).


\begin{figure}[htbp]
\centering

 \begin{minipage}{.39\textwidth}
  \centering
  \includegraphics[width=.99\linewidth]{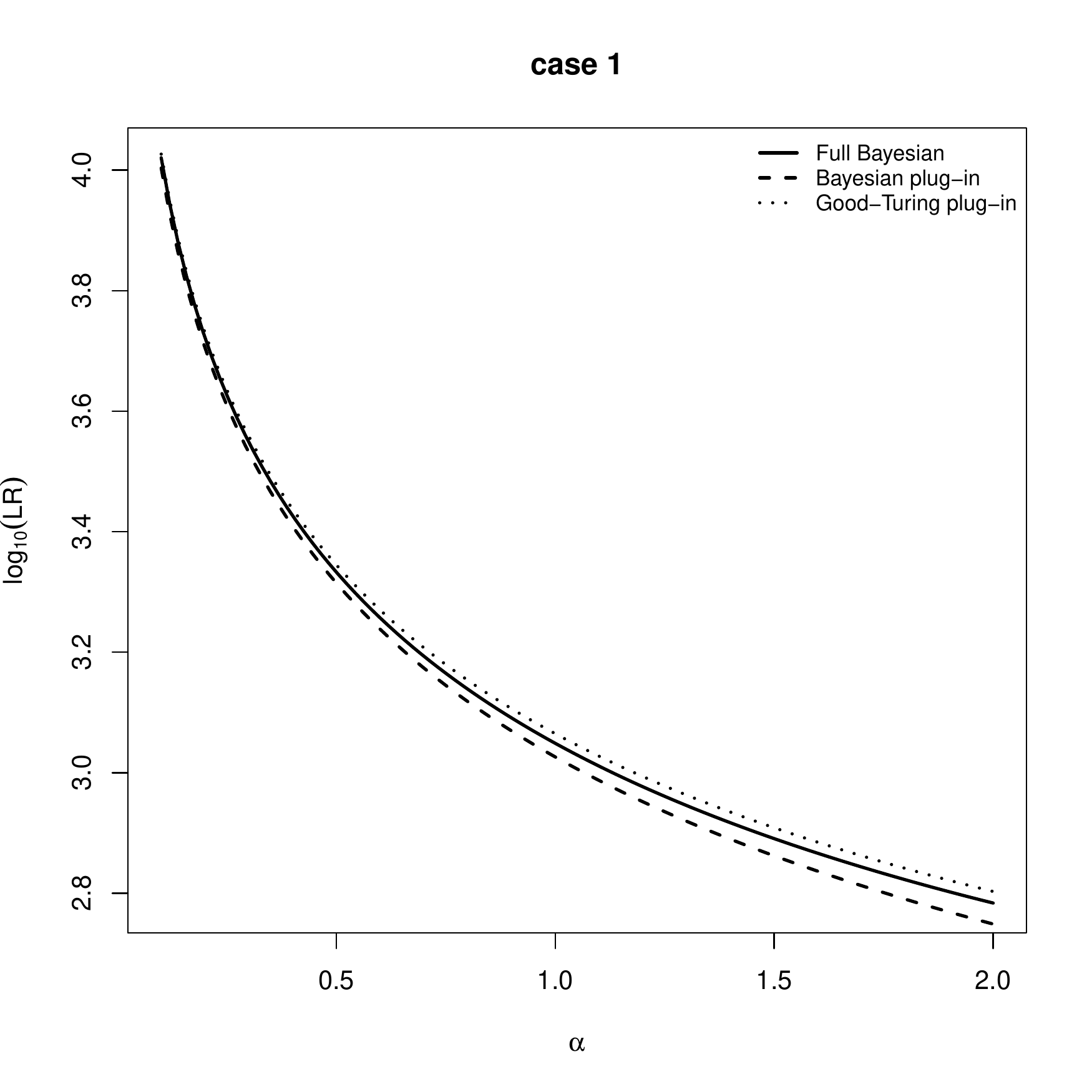}
\end{minipage}
\begin{minipage}{.39\textwidth}
  \includegraphics[width=.99\linewidth]{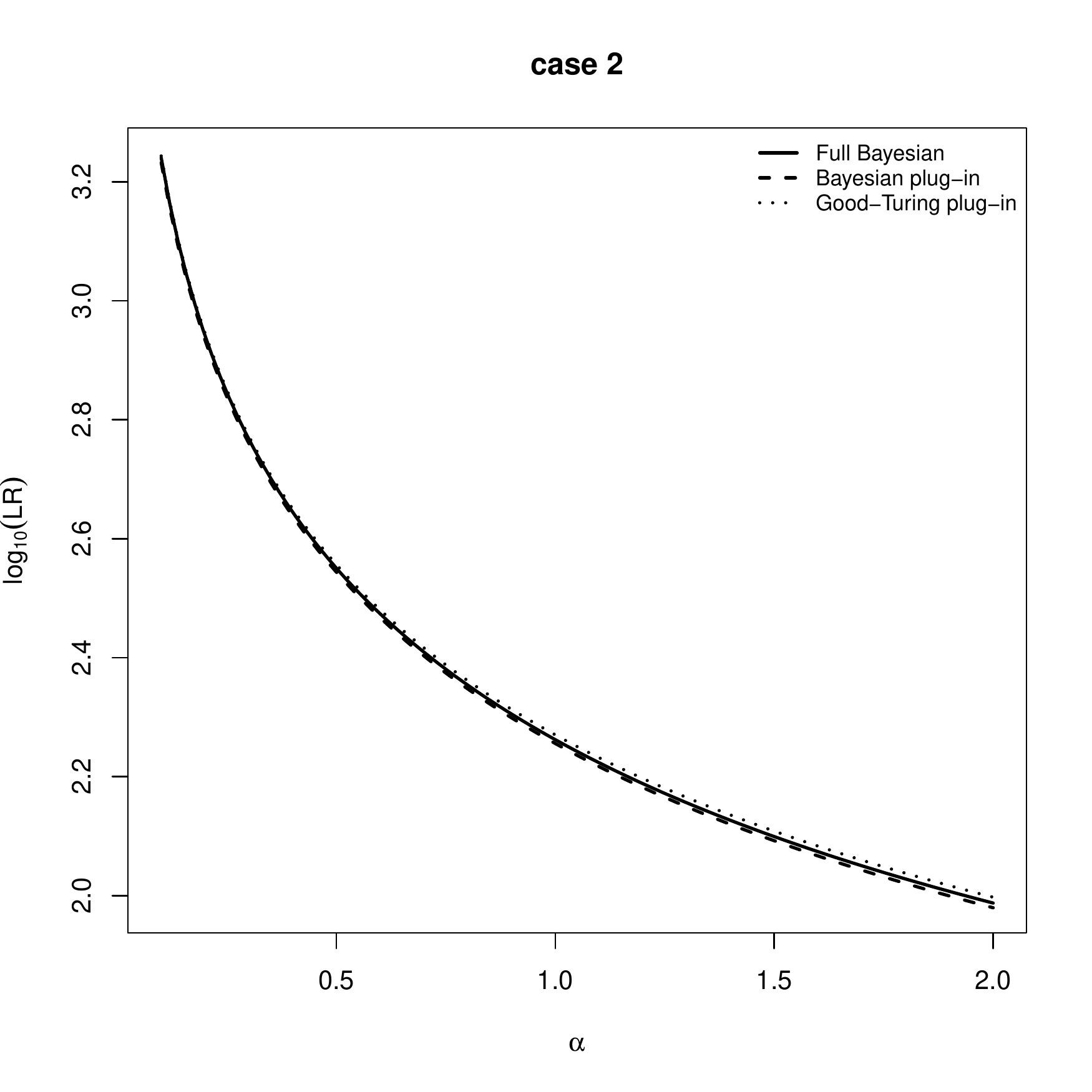}

  \centering
\end{minipage}
\caption{\small Sensitivity analysis for the $\log_{10}(\LR)$ obtained with (i) the full Bayesian approach, (ii) the hybrid Good-Turing plug-in (iii) classical Bayesian plug-in, for the two cases described in Table~\ref{acsesd} when the prior over $k^L$ and $k^S$ is uniform over $\{1,..., m\}$, \textcolor{black}{with $\alpha=1$ and $m=100$.}}  \label{ddefault}

\end{figure}

The sensitivity analysis for the $\log_{10}(\LR)$, shown in Figure~\ref{ddefault}, has been conducted for different loci and different combinations of alleles, without showing substantial differences (in terms of sensitivity). Moreover, it tells us that the two plug-in approaches represent acceptable solutions in terms of quantification.
\textcolor{black}{We carried on additional experiments that showed us that varying $m$ does not change Figure~\ref{ddefault} in a way which is significative for the case at hand.}


\section{Conclusion}\label{sdsdf}
Mostly due to the limited size of the available database (about one hundred people in a given relevant population), the rare type match situation is very likely to be encountered when DIP-STR data is used. The recipients of this new technology should be prepared for such an eventuality, which was not taken into account in the OOBN proposed in \citet{cereda:2014}.
This paper provides a methodology that allows one to obtain the full Bayesian likelihood ratio also when there are DIP-STR alleles which are not present in the reference database among the alleles of the known contributor and of the suspect. This is done by extending the OOBN, and introducing a more complex prior over the allelic frequencies (a mixture of Dirichlet and uniform distribution) based on a previously developed solution for Y-STR data \citep{cereda:2015}.
Notice that this issue also represents an opportunity to discuss the use of plug-in approximations which are compared with the full Bayesian likelihood ratio.
They proved to be valid approximations.

The sensitivity analysis of the hyperparameters of the prior is also studied. The results show that the likelihood ratio moderately depends on the choices of the parameters $\alpha$ of the Dirichlet prior. 
Hence, there is the need for further investigations to find better priors, either less sensitive to hyperparameters, or more realistic, such as it was done for Y-STR data in \citet{cereda:2015c}. \textcolor{black}{Alternatively, we can hope in more data to make the choice of the prior less important. By looking at Figure \ref{ddefault}, we can conclude that unless more data are available, the LR can be determined up to an order of magnitude at best.}

\section*{Appendix. Full Bayesian likelihood ratio development}\label{appendix}

\begin{table}[htbp]
\begin{center}
\begin{tabular}{|p{1cm} |  p{8.8cm} |c|}
\hline
Name& Description&  Type\\
\hline
$m$				& number of theoretically possible L-STR (and S-STR) alleles							& fixed	\\

			
$k^{L}$				& number of L-STR allele types present in the population 						& random \\
$k^{S}$						& number of S-STR allele types present in the population 						& random \\
$\mathbf{t}^L$				& positions (from $1$ to $m$) of the $k^{L}$ L-STR allele types in the population	& random \\
$\mathbf{t}^S$		 		& positions (from $1$ to $m$) of the $k^{S}$ S-STR allele types in the population	& random \\

$\boldsymbol{\theta}$		& population proportions of the $2m$ possible DIP-STR alleles 	& random \\
$\psi$		 		& sum of the relative frequencies of the L-STR alleles & random  \\
$\boldsymbol{\phi}^L$		 		& normalised vector of the relative frequencies of L-STR alleles & random  \\
$\boldsymbol{\phi}^S$		 		& normalised vector of the relative frequencies of S-STR alleles & random  \\

$n$				& size of the available database								& observed\\
$n^{L}$			 		& total number of L-STR alleles in the augmented database			& observed\\
$n^{S}$			 		& total number of S-STR alleles in the augmented database			& observed\\
$\mathbf{b}$		 	 		& labels $(j,i)$ corresponding to each of the $n+4$ DIP-STR alleles in the augmented database	& observed \\

$\mathbf{b}^L$		 	 		& labels $(\textrm{L}, i)$ corresponding to each of the $n^L$ L-STR alleles in the augmented database	& observed \\

$\mathbf{b}^S$		  	& labels $(\textrm{S}, i)$ corresponding to each of the $n^S$ S-STR alleles in the augmented database	& observed \\

$k^L_{\mathbf{b}}$					& number of distinct  L-STR alleles in the augmented database		& observed\\
$k^S_{\mathbf{b}}$				& number of distinct  S-STR alleles in the augmented database		& observed\\
$\mathbf{n}^L$			 		& counts of all $m$ L-STR alleles in the augmented database	& observed \\
$\mathbf{n}^S$		 		& counts of all $m$ S-STR alleles in the augmented database	& observed \\

\hline

\end{tabular}
\caption{\small Some relevant symbols used in the paper.}\label{Table4w44}
\end{center}
\end{table}

In Table~\ref{Table4w44}, a summary of the relevant symbols used is reported. 
The aim of this Appendix is to develop the conditional expectation of the functions reported in Table~\ref{Table2ed}, which constitute the denominator of the likelihood ratio \eqref{eq1e}. Those conditional expectations can be rewritten in terms of $\boldsymbol{\phi}^L$, $\boldsymbol{\phi}^S$, and $\psi$, in the following way:

$$\mathbb{E}(2\Theta^L_{i}\Theta^L_j\mid \mathbf{b})=\mathbb{E}(2\Phi^L_i\Phi^L_j\Psi^2\mid \mathbf{b})=2\mathbb{E}(\Phi^L_i\Phi^L_j\mid \mathbf{b})\mathbb{E}(\Psi^2\mid \mathbf{b}),$$
 $$\mathbb{E}((\Theta^{L}_{i})^2+2 \Theta^L_i (1-\Psi) \mid \mathbf{b})=\mathbb{E}((\Phi^{L}_{i})^2\mid \mathbf{b})\mathbb{E}(\Psi^2\mid \mathbf{b}) + 2\mathbb{E}( \Phi^L_i \mid \mathbf{b})\mathbb{E}( \Psi \mid \mathbf{b})\mathbb{E}(1- \Psi \mid \mathbf{b}).$$

\subsection*{The distribution of $\Psi$ given $\mathbf{B}$.}
As explained in Section~\ref{sol}, $\boldsymbol{\theta}$ can be represented through a set of three independent variables ($\boldsymbol{\phi}^L$, $\boldsymbol{\phi}^S$, $\psi$). The vector $\mathbf{b}$ can also be reduced by sufficiency to three random variables: $(n^{L}, \mathbf{n}^L, \mathbf{n}^S)$, where $n^{L}$ is the total number of observed L-STR alleles in the enlarged database, $\mathbf{n}^L$ is the vector of length $m$ containing the counts in the augmented database of each of the $m$ L-STR alleles, in an order that corresponds to that of $\boldsymbol{\phi}^L$, $\mathbf{n}^S$ is the vector of counts of each of the $m$ S-STR alleles. $n^{L}$ is binomial distributed with parameters ($n+4$, $\boldsymbol{\psi}$), while $\mathbf{n}^L$ is multinomial distributed with parameters ($n^{L}, \boldsymbol{\phi}^L$). Similarly, $\mathbf{n}^S$ is multinomial distributed with parameters ($n^{S}, \boldsymbol{\phi}^S$), where $n^S=n+4-n^L$ is the number of S-STR alleles in the augmented database.

It holds that the likelihood for $\boldsymbol{\phi}^L, \boldsymbol{\phi}^S$, and $\psi$ factors:
$$p(n^L, \mathbf{n}^L,\mathbf{n}^S \mid \boldsymbol{\phi}^L, \boldsymbol{\phi}^S, \psi)= p(n^{L}\mid \psi) p(\mathbf{n}^L\mid n^{L}, \boldsymbol{\phi}^L)p(\mathbf{n}^S\mid n^{S}, \boldsymbol{\phi}^S).
$$
The priors for $\boldsymbol{\phi}^L, \boldsymbol{\phi}^S$, and $\psi$ factors as well, since they are independent. Therefore, the posteriors for $\boldsymbol{\phi}^L, \boldsymbol{\phi}^S$, and for $\psi$ given $\mathbf{b}$ factors as the product of three independent posteriors. Thus, it holds that
$$
p(\psi \mid \mathbf{b})\propto p(n^{L}\mid \psi) p(\psi),
$$
which is a product of the density of a binomial distribution and of a beta prior. By conjugacy,
$$\Psi \mid \mathbf{B}=\mathbf{b}\sim \textrm{Beta}(1+ n^{L}, 1+n^S).$$

In conclusion, by using properties of the Beta distribution, it holds that 
\begin{equation}\label{giu1}
\mathbb{E}(\Psi^2\mid \mathbf{b} )= \frac{(n^{L}+1)(n^{L}+2)}{(n+6)(n+7)},
\end{equation}
and 
\begin{equation}\label{giu1b}
\mathbb{E}((1-\Psi)^2\mid \mathbf{b} )=\frac{(n^{S}+1)(n^{S}+2)}{(n+6)(n+7)}.
\end{equation}

\subsection*{The distribution of $\boldsymbol{\phi}^L$ and $\boldsymbol{\phi}^S$ given $\mathbf{B}$.}
Let $p(k)$ be the prior distribution over $k^{L}$ and $k^{S}$. 
In this section we will omit superscripts L and S from $k$, $\mathbf{t}$, $\boldsymbol{\phi}$, and $\mathbf{n}$, in order to obtain general results valid for both cases. Notice that $n$ will stand for $n^L$ or $n^S$, and $\mathbf{b}$ will stand for $\mathbf{b}^L$ or $\mathbf{b}^S$ as described in Table~\ref{Table4w44} (so temporarily, the meaning of $n$, and $\mathbf{b}$ is different from its meaning in the rest of the paper).

Given $k $, $\mathbf{t}$ is uniformly distributed over the ordered vectors containing $k$ indexes from 1 to $m$.
Let us denote with $k_{\mathbf{b}}$ the number of distinct L-STR (or S-STR) alleles observed in the augmented database, and with $\boldsymbol{{\phi}_{\mathbf{b}}}$ the vector of length $k_{\mathbf{b}}$ containing only the frequencies of the L-STR alleles observed in the augmented database in the order in which they appear in $\boldsymbol{\phi}$. $\boldsymbol{\phi_{\mathbf{b}}}$ does not sum to one, since there are L-STR alleles of positive frequency, which are not observed: the total probability mass of the unobserved alleles is $\bar{\phi}_{\mathbf{b}}=1-\sum_{i=1}^{k_{\mathbf{b}}}{\phi_{\mathbf{b}i}}$. The vector $\boldsymbol{\phi_{\mathbf{b}}}^*=(\boldsymbol{\phi_{\mathbf{b}}}, \bar{\phi}_{\mathbf{b}})$ sums up to one.

We can look for the posterior distribution of $\boldsymbol{\phi_{\mathbf{b}}}^*$ given the vector $\mathbf{b}$.  

\begin{equation}\label{eqwsee}
p(\boldsymbol{\phi_{\mathbf{b}}}^*\mid \mathbf{b})= \sum_{k}\sum_{\mathbf{t}}p(\boldsymbol{\phi_{\mathbf{b}}}^*\mid \mathbf{b},\mathbf{t} )p(\mathbf{t}\mid k, \mathbf{b})p(k\mid \mathbf{b})
\end{equation}

It can be proved that 
\begin{itemize}
\item the posterior density $p(\boldsymbol{\phi_{\mathbf{b}}}^*\mid \mathbf{b},\mathbf{t})$ depends on $\mathbf{t}$ only through $k$. Hence, we can denote it as $p(\boldsymbol{\phi_{\mathbf{b}}}^*\mid \mathbf{b}, k)$
\item if $k$ is less than $k_{\mathbf{b}}$, then $p(k\mid \mathbf{b})=0$.
\item let us denote with $\mathcal{T}_{k, \mathbf{b}}$ the set of ordered vectors $\mathbf{t}$ of length $k$ and compatible with $\mathbf{b}$ (i.e., which contain among others the positions corresponding to the elements in $\mathbf{b}$).
For all the $\mathbf{t}$ which are not in $\mathcal{T}_{k,  \mathbf{b}}$, then $p(\mathbf{t}\mid k, \mathbf{b})=0$.
\end{itemize}

We can change the summation indexes in \eqref{eqwsee} to obtain:
$$
p(\boldsymbol{\phi_{\mathbf{b}}}^* \mid \mathbf{b})=\sum_{k= k_{\mathbf{b}}}^{m} p(k\mid \mathbf{b})p(\boldsymbol{\phi_{\mathbf{b}}}^*  \mid k, \mathbf{b}) \sum_{\mathbf{t} \in \mathcal{T}_{k,  \mathbf{b}} }p( \mathbf{t} \mid k, \mathbf{b}).$$

For any of the $\binom{m-k_{\mathbf{b}}}{k-k_{\mathbf{b}}}$ vectors $\mathbf{t}$ in $\mathcal{T}_{k, \mathbf{b}}$, $p(\mathbf{t}\mid k, \mathbf{b})$ has the same value $\frac{1}{\binom{m-k_{\mathbf{b}} }{k-k_{\mathbf{b}}}}$.
Thus, in the end we have that
$$p(\boldsymbol{\phi_{\mathbf{b}}}^* \mid \mathbf{b})=\sum_{k= k_\mathbf{b}}^{m} p(k\mid \mathbf{b})p(\boldsymbol{\phi_{\mathbf{b}}}^* \mid k, \mathbf{b}).$$
The distribution $p(k\mid \mathbf{b})$ can be obtained in the following way. 
$$p(k, \mathbf{t}, \boldsymbol{\phi}, \mathbf{b})= p(k) p(\mathbf{t}\mid k) p(\boldsymbol{\phi}\mid \mathbf{t})p(\mathbf{b}\mid \boldsymbol{\phi}).$$
Integrating out $\boldsymbol{\phi}$, we obtain
\begin{equation}\label{4t}
p(k, \mathbf{b}, \mathbf{t})= p(k) p(\mathbf{t}\mid k) \int_{\boldsymbol{\phi}}p(\boldsymbol{\phi}\mid \mathbf{t})p(\mathbf{b}\mid \boldsymbol{\phi})\mathrm{d}\boldsymbol{\phi},
\end{equation}
where the integral contains a Dirichlet density and the categorical density defined in \eqref{fed}. They are conjugate, thus we obtain
$$
p(k, \mathbf{t}\mid \mathbf{b})\propto p(k) p(\mathbf{t}\mid k) \frac{\Gamma(k\alpha)}{\Gamma(n+k\alpha)}.	
$$
Now we can sum over the $\mathbf{t}$ compatible with $\mathbf{b}$, to get to 
\begin{equation}
p(k\mid  \mathbf{b})\propto 	\binom{k}{k_{\mathbf{b}}}p(k) \frac{\Gamma(k\alpha)}{\Gamma(n+k\alpha)}.
\end{equation}

In conclusion, %
\begin{equation}\label{erd}
p(\boldsymbol{\phi_{\mathbf{b}}}^* \mid \mathbf{b}) \propto \sum_{k= k_{\mathbf{b}}}^{m} \binom{k}{k_{\mathbf{b}}}p(k) \frac{\Gamma(k\alpha)}{\Gamma(n+k\alpha)}	p(\boldsymbol{\phi_{\mathbf{b}}}^* \mid k, \mathbf{b}),
\end{equation}
where $\boldsymbol{\Phi_{\mathbf{b}}}^* \mid  K=k, \mathbf{B}=\mathbf{b} \sim \text{Dir}^{k_{\mathbf{b}}+1}(\alpha+\tilde{n}_1, ..., \alpha+\tilde{n}_{k_{\mathbf{b}}},(k-k_{\mathbf{b}})\alpha )$, and $\mathbf{\tilde{n}}$ is the vector of length $k_{\mathbf{b}}$ with the positive elements of $\mathbf{n}$.

Therefore, \eqref{erd} is a mixture of Dirichlet distributions with weights $w(k)=\binom{k}{k_{\mathbf{b}}}p(k) \frac{\Gamma(k\alpha)}{\Gamma(n+k\alpha)}.$
Using properties of the Dirichlet distribution we obtain that, $\forall i,j $ corresponding to different observed DIP-STR alleles: 
\begin{equation}\label{giu2}
\mathbb{E}(\Phi_{i}\Phi_{j}\mid \mathbf{b}  )=(\alpha+{n_{i}})(\alpha+{n_{j}}) \frac{\sum_{k= k_{\mathbf{b}}}^{m} w(k) g(k) }{\sum_{k= k_{\mathbf{b}}}^{m} w(k)},
\end{equation}

\begin{equation}\label{giu3}
\mathbb{E}(\Phi_{i}^2\mid \mathbf{b} )=(\alpha+n_{i})(\alpha+n_{i}+1) \frac{\sum_{k= k_{\mathbf{b}}}^{m} w(k) g(k) }{\sum_{k= k_{\mathbf{b}}}^{m} w(k)},
\end{equation}
where  $g(k)=\frac{1}{(k\alpha+n)(k\alpha+n+1)}$.

\subsection*{The conditional expectations in Table~\ref{Table2ed}}

Using \eqref{giu1}, \eqref{giu1b}, \eqref{giu2} and \eqref{giu3}, we obtain that, $\forall i, j$ corresponding to different observed alleles,
 \begin{align}\label{ess2}
\mathbb{E}(\Theta^L_i\Theta^L_j \mid \mathbf{b} )&=\mathbb{E}(\Phi^{L}_i\Phi^{L}_j \Psi^2 \mid \mathbf{b})=\mathbb{E}(\Phi^{L}_i\Phi^{L}_j\mid \mathbf{b}  )\mathbb{E}(\Psi^2 \mid \mathbf{b})\\&=(\alpha+{n^L_{i}})(\alpha+{n^{L}_j}) \frac{\sum_{k= k^L_{\mathbf{b}}}^{m} w^L(k) g^L(k) }{\sum_{k= k^L_{\mathbf{b}}}^{m} w^L(k)}\frac{(n^{L}+1)(n^{L}+2)}{(n+6)(n+7)}, \end{align}

\begin{align}\label{ess23}
\mathbb{E}((\Theta_i^L)^2 \mid \mathbf{b} )=&\mathbb{E}((\Phi^L_i)^2\Psi^2 \mid \mathbf{b})=\mathbb{E}((\Phi^L_i)^2\mid \mathbf{b}  )\mathbb{E}(\Psi^2 \mid \mathbf{b})=\\
=&(\alpha+n_{i}^L)(\alpha+n_{i}^L+1)\frac{\sum_{k= k^L_{\mathbf{b}}}^{m} w^L(k) g^L(k) }{\sum_{k= k^L_{\mathbf{b}}}^{m} w^L(k)}\frac{(n^{L}+1)(n^{L}+2)}{(n+6)(n+7)}, \end{align}

\begin{align}\label{ess23}
\mathbb{E}(\Theta_i^L(1-\Psi) \mid \mathbf{b} )=&\mathbb{E}(\Phi^L_i\mid \mathbf{b})(\mathbb{E}(\Psi \mid \mathbf{b})-\mathbb{E}(\Psi^2 \mid \mathbf{b}))=\\
=&(\alpha+n^L_{i})\frac{\sum_{k= k^L_{\mathbf{b}}}^{m} \frac{w^L(k)}{k\alpha + n^L} }{\sum_{k= k^L_{\mathbf{b}}}^{m} w^L(k)}\frac{(n^L+1)(n^S+1)}{(n+6)(n+7)}, \end{align}
where $n$ and $\mathbf{b}$ have now their original meaning, and $w^L(k)=\binom{k}{k^L_{\mathbf{b}}}p(k) \frac{\Gamma(k\alpha)}{\Gamma(n^L+k\alpha)}$, and $g^L(k)=\frac{1}{(k\alpha+n^L)(k\alpha+n^L+1)}$.

These formulas can be directly used to obtain the conditional expectations of the last three rows of Table~\ref{Table2ed}. In a very similar way one can easily obtain the conditional expectations contained in the first three rows.

\bibliographystyle{apalike} 

\end{document}